\documentclass[prd,superscriptaddress,showpacs,twocolumn,nofootinbib,floatfix,amsmath,amssymb]{revtex4-2}
\usepackage{mathtools}
\usepackage{mathrsfs}
\usepackage{color}
\usepackage{braket}
\usepackage{bm}
\usepackage{graphicx}
\usepackage{hyperref}
\usepackage{epstopdf}

\newcommand{\reals}{\mathbb{R}}

\DeclareMathOperator{\Tr}{Tr}

\newcommand\identity{1\kern-0.25em\text{l}}

\begin{document}

\title{Reply to `Comment on ``Ideal clocks -- a convenient fiction'''}

\author{Krzysztof Lorek}\thanks{klorek1@gmail.com} 

\affiliation{XIV Stanis\l{}aw Staszic High School, Nowowiejska 37a, 02-010 Warsaw, Poland}

\author{Jorma Louko}\thanks{jorma.louko@nottingham.ac.uk}

\affiliation{School of Mathematical Sciences, University of Nottingham, Nottingham NG7 2RD, United Kingdom}

\author{Andrzej Dragan}\thanks{dragan@fuw.edu.pl}

\affiliation{Institute of Theoretical Physics, University of Warsaw, Pasteura 5, 02-093 Warsaw, Poland}
\affiliation{Centre for Quantum Technologies, National University of Singapore, 3 Science Drive 2, 117543 Singapore, Singapore}

\date{April 2026, revised August 2026.\\aa Published in Class.\ Quantum Grav.\ \textbf{43}, 158001 (2026), doi.org/10.1088/1361-6382/ae8e23}

\begin{abstract}
For a quantum scalar field that is confined in a uniformly linearly accelerated cavity in Minkowski spacetime and interacts linearly with a scalar field that is not confined in the cavity, a de-excitation probability formula was obtained in 
Lorek \emph{et al\/} [Class.\ Quantum Grav.\ \textbf{32}, 175003 (2015)] 
by a first-order perturbation theory calculation. A~recent Comment by Toussaint 
[Class.\ Quantum Grav.\ \textbf{43}, 068001 (2026)] 
questions this formula on the grounds that the calculation in Lorek \emph{et al\/} invokes Rindler modes both in the Rindler wedge of the accelerated cavity and in the opposing, causally disconnected Rindler wedge. In the present Reply we rederive the de-excitation formula given in Lorek \emph{et al\/} by a perturbation theory calculation that is formulated entirely within the Rindler wedge of the accelerated cavity. 
We also take the opportunity to comment on the role of the two sets of Rindler modes in the calculation presented in Lorek \emph{et al\/}. 
\end{abstract}

\maketitle

\section{Introduction}

For a quantum scalar field that is confined in a uniformly linearly accelerated cavity in Minkowski spacetime and interacts linearly with a scalar field that is not confined in the cavity, a de-excitation probability formula was obtained in 
\cite{Lorek:2015rua} by a first-order perturbation theory calculation. 
A recent Comment \cite{Toussaint:2026jop}
questions this formula on the grounds that the calculation in \cite{Lorek:2015rua} invokes in its intermediate stages Rindler modes both in the Rindler wedge of the accelerated cavity and in the opposing, causally disconnected Rindler wedge. 

In the present Reply we rederive the de-excitation formula given in \cite{Lorek:2015rua} by a perturbation theory calculation that is formulated entirely within the Rindler wedge of the accelerated cavity, 
using the fact that this wedge is a globally hyperbolic spacetime in its own right, and evolving in time in a Cauchy foliation in this wedge. 
We also take the opportunity to comment on the role of the two sets of Rindler modes in the calculation presented in~\cite{Lorek:2015rua}. 

\section{Uncoupled system}

In the notation of~\cite{Lorek:2015rua}, we work in $(1+1)$-dimensional Minkowski spacetime, with the metric $ds^2 = - dt^2 + dx^2$, and therein in the right-hand-side Rindler wedge $x>|t|$. We use the Rindler coordinates $(\tau,\xi)$, given by 
\begin{subequations}
\begin{align}
t &= \alpha^{-1} e^{\alpha \xi} \sinh(\alpha \tau) \,,
\\
x &= \alpha^{-1} e^{\alpha \xi} \cosh(\alpha \tau) \,,
\end{align}
\end{subequations} 
where $\alpha$ is a positive parameter of dimension inverse length. 
The metric reads 
\begin{align}
ds^2 = e^{2\alpha\xi} \! \left( - d\tau^2 + d\xi^2 \right) \,,
\end{align}
with $\tau\in\reals$ and $\xi\in\reals$. 
The worldlines of constant $\xi$ are uniformly accelerated, 
with proper acceleration~$\alpha e^{-\alpha\xi}$, and $\tau$ is the proper time on the worldline $\xi=0$. 

In the Rindler wedge, we consider a cavity of proper length $l>0$, at $\xi_- \le \xi \le \xi_+$, 
where $\xi_\pm = \alpha^{-1} \ln(\alpha\sigma_\pm)$ and $\sigma_\pm = \alpha^{-1} \pm l/2$. The centre of the cavity is at $\xi=0$. 

In the cavity, we consider a quantised real massless scalar field $\phi$ with Dirichlet boundary conditions. $\phi$~has the mode expansion 
\begin{align}
\phi = \sum_{n=1}^\infty  \bigl( b_n u_n + b_n^\dagger u_n^* \bigr) \,,
\end{align}
where $u_n$ are a set of Klein-Gordon orthonormal mode functions of positive frequency 
$\omega_n = \alpha n \pi / \ln(\sigma_+/\sigma_-)$ 
with respect to~$\partial_\tau$, and the nonvanishing commutators of the creation and annihilation operators are $[b_n, b_{n'}^\dagger] = \delta_{n n'}$. The explicit formula for $u_n$ is given by (21) in~\cite{Lorek:2015rua}. 

In the full Rindler wedge, we consider an ambient quantised real scalar field $\Phi$ of mass $M>0$. $\Phi$~has the mode expansion 
\begin{align}
\Phi = \int_0^\infty \text{d}\Omega \, \bigl( B_\Omega U_\Omega + B_\Omega^\dagger U_\Omega^* \bigr) \,, 
\end{align}
where 
$U_\Omega$ are a set of Klein-Gordon continuum-orthonormal mode functions of positive frequency $\Omega$ with respect to~$\partial_\tau$, and the nonvanishing commutators of the creation and annihilation operators are $[B_\Omega, B_{\Omega'}^\dagger] = \delta(\Omega - \Omega')$. 
The explicit formula for $U_\Omega$ is given by (22) and (23) in~\cite{Lorek:2015rua}. 
We have here dropped the Rindler wedge index that appears in \cite{Lorek:2015rua} since here we are throughout in the right-hand-side Rindler wedge.

\section{Initial state}

We prepare the cavity field $\phi$ in the state where the lowest cavity mode ($n=1$) is in its first excited Fock state. We denote this state by $|1\rangle_1$, and the corresponding density matrix by $\rho_{\phi,i} = |1\rangle_1 \, {}_1\langle 1|$. 

We prepare the ambient field $\Phi$ in the mixed state that is induced in the Rindler wedge by the Minkowski vacuum. In our notation, this state has the density matrix \cite{Birrell1984}
\begin{align}
\rho_{\Phi,i} = \bigotimes_\Omega  \bigl( 1 - e^{-2\pi\Omega/\alpha}\bigr) \sum_{n=0}^\infty e^{-2\pi n \Omega/\alpha} \, |n\rangle_\Omega \, {}_\Omega\langle n| \,, 
\label{eq:rhoPhi-initial}
\end{align}
where $|n\rangle_\Omega$ is the $n$-particle Fock state in the field mode~$\Omega$. 

The initial density matrix of the total system is $\rho_i = \rho_{\phi,i} \otimes \rho_{\Phi,i}$.

\section{Coupled evolution}

Recall that the Rindler wedge is a globally hyperbolic spacetime, and the hypersurfaces of constant $\tau$ provide a Cauchy foliation. We use this foliation to evolve the coupled system in time. 

We take the interaction Hamiltonian with respect to the evolution in $\tau$ to be 
\begin{align} 
\label{HInt}
H_{\text{int}}(\tau) = \lambda\int_{\xi_-}^{\xi_+}\text{d}\xi\,\phi(\tau,\xi) \Phi(\tau,\xi) \,, 
\end{align}
where $\lambda$ is a real-valued coupling constant. 
We work in the interaction picture, perturbatively in~$\lambda$. 

After evolving the coupled system from $\tau = \tau_i$ to $\tau = \tau_f$, the density matrix becomes  
\begin{align} 
\rho_f = U \rho_i U^{\dagger} \,,
\end{align}
where 
\begin{subequations}
\begin{align} 
U &= \identity + U_1 + U_2 + O(\lambda^3) \,,
\\
U_1 &= - i \lambda \int_{\tau_i}^{\tau_f} \text{d}\tau_1 \, H_{\text{int}}(\tau_1) \,,
\label{eq:U1-def}
\\
U_2 &= - \lambda^2 \int_{\tau_i}^{\tau_f} \text{d}\tau_1 \, H_{\text{int}}(\tau_1) 
\int_{\tau_i}^{\tau_1} \text{d}\tau_2 \, H_{\text{int}}(\tau_2) 
\,, 
\end{align}
and $\identity$ denotes the identity operator on the total Hilbert space. 
\end{subequations}
Hence 
\begin{subequations}
\label{eq:rhof-decomp}
\begin{align} 
\rho_f & = \rho_i + \rho_f^{(1)} + \rho_f^{(2)} + O(\lambda^3) \,,
\\
\rho_f^{(1)} &= U_1 \rho_i + \rho_i U_1^\dagger \,,
\\
\rho_f^{(2)} &= U_1 \rho_i U_1^\dagger + U_2 \rho_i + \rho_i U_2^\dagger \,.
\label{eq:rhof-decomp2}
\end{align}
\end{subequations}

\section{Cavity de-excitation probability}

The probability to find the cavity field $\phi$ in its ground state $|0\rangle_\phi$ at $\tau = \tau_f$ is 
\begin{align}
P_{\downarrow} = \Tr \! \left( ( |0\rangle_\phi \, {}_\phi \langle0| \otimes \identity_\Phi ) \, {\rho_f}_{\vphantom{A}} \right) \,, 
\end{align}
where $\identity_\Phi$ denotes the identity operator on the Hilbert space of~$\Phi$, 
and the trace is over the total Hilbert space. 
From \eqref{eq:rhof-decomp} it is seen that the leading contribution to $P_{\downarrow}$ 
comes from the term $U_1 \rho_i U_1^\dagger$ in $\rho_f^{(2)}$ \eqref{eq:rhof-decomp2} and is of order~$\lambda^2$. 
Writing $P_{\downarrow} = P_{\downarrow}^{(2)} + O(\lambda^3)$, 
\eqref{eq:rhoPhi-initial} and \eqref{eq:U1-def} hence give 
\begin{align}
P_{\downarrow}^{(2)} 
&= 
\lambda^2 \int_0^\infty \text{d}\Omega \, \bigl( 1 - e^{-2\pi\Omega/\alpha}\bigr) 
\notag
\\
& \hspace{7ex}
\times 
\sum_{n=0}^\infty e^{-2\pi n\Omega/\alpha}
\left( (n+1) |\gamma_{\Omega 1}|^2  + n |\tilde \gamma_{\Omega 1}|^2 \right)
\notag
\\
&= 
\lambda^2 \int_0^\infty 
\frac{\text{d}\Omega} {1 - e^{-2\pi\Omega/\alpha}} 
\left( 
|\gamma_{\Omega 1}|^2 
+ e^{-2\pi\Omega/\alpha} |\tilde \gamma_{\Omega 1}|^2 
\right) 
\notag
\\
&= 
\lambda^2 \int_0^\infty 
\text{d}\Omega 
\left( 
\cosh^2 \! r_\Omega \, |\gamma_{\Omega 1}|^2 + \sinh^2 \! r_\Omega \, |\tilde \gamma_{\Omega 1}|^2 
\right) \,,
\label{eq:Pdown}
\end{align}
where 
\begin{subequations}
\begin{align}
\gamma_{\Omega n} & = \int_{\tau_i}^{\tau_f}\text{d}\tau\int_{\xi_-}^{\xi_+}\text{d}\xi\,U_\Omega^*(\tau, \xi) u_n(\tau,\xi) \,,
\\
\tilde\gamma_{\Omega n} & = \int_{\tau_i}^{\tau_f}\text{d}\tau\int_{\xi_-}^{\xi_+}\text{d}\xi\,U_\Omega(\tau, \xi) u_n (\tau, \xi) \,,
\end{align}
\end{subequations}
and 
\begin{eqnarray}
\tanh r_\Omega = e^{-\pi\Omega/\alpha} \,. 
\end{eqnarray}

Equation \eqref{eq:Pdown} agrees with equation (19) in~\cite{Lorek:2015rua}. This is what we set out to demonstrate. 

\section{Discussion}

In~\cite{Lorek:2015rua}, the de-excitation probability formula \eqref{eq:Pdown} was obtained by a calculation in which the evolution is done in the Rindler time of the Rindler wedge of the cavity, but the initial state for the ambient field $\Phi$ is written as the Minkowski vacuum on full Minkowski spacetime. The intermediate steps in the calculation hence involve two sets of Rindler modes, one set in the Rindler wedge of the cavity and the other in the opposing, causally disconnected wedge. We now briefly discuss why the inclusion of both sets of Rindler modes in the intermediate steps does not contradict causality of the evolution in the cavity. 

To set the stage, consider the corresponding de-excitation calculation for an inertial cavity, as done in~\cite{Lorek:2015rua}. The intermediate steps of the calculation involve Minkowski plane wave modes, which have support in parts of Minkowski space that are spacelike separated from the part of the cavity's trajectory where the interaction occurs. This however does not contradict causality: causality is enforced by the support of the interaction. 

Consider then the accelerating cavity, evolved in Rindler time of the cavity's Rindler wedge as in~\cite{Lorek:2015rua}. As the interaction is nonzero only within the cavity, the evolution within the cavity is independent of how one chooses to continue the constant time slices outside the cavity. We may hence envisage the time slices outside the cavity to be deformed for example so that they are asymptotic to hypersurfaces of constant Minkowski time in the far right and far left. The causality considerations then become similar to those for the inertial cavity. The intermediate steps in the calculation again involve Minkowski one-particle states, whose mode functions are spread as plane waves over all of the spacetime, including regions that are spacelike separated from the part of the cavity's trajectory where the interaction occurs. When the intermediate Minkowski one-particle mode functions are written in terms of the Rindler mode functions, they involve Rindler mode functions in both of the two Rindler wedges; this however does not contradict causality of the evolution within the cavity, 
which is again enforced by the support of the interaction. 

\section*{Acknowledgements}

We thank Vladimir Toussaint for correspondence.

\end{document}